\documentclass[5p,times,authoryear,nopreprintline]{elsarticle}

\usepackage[authoryear,round]{natbib}
\usepackage{graphicx}
\usepackage{caption}
\usepackage{subcaption}
\usepackage{amsmath}
\usepackage{amssymb}
\usepackage{comment}
\usepackage{lipsum}
\newcommand{\arcsec}{arcsec}

\usepackage{tikz}
\usepackage{tikzscale}
\usepackage{pgfplots}
\pgfplotsset{compat=1.18}

\usepackage{url}
\usepackage[colorlinks=true,linkcolor=blue,citecolor=blue,urlcolor=blue]{hyperref}

\usepackage{url}

\journal{Astronomy and Computing}

\begin{document}

\begin{frontmatter}

\title{\vspace{-2.5em}{\normalfont\footnotesize\textbf{Accepted for publication in Astronomy and Computing (January 2027)} \\
\footnotesize DOI: \href{https://doi.org/10.1016/j.ascom.2026.101198}{10.1016/j.ascom.2026.101198}\\[1.8em]}%
LUMA: A CNN for Strong Gravitational Lens Searches in Astronomical Imaging}

\author[aff1,aff2]{Giovanni Vincenzo Donatiello\corref{cor1}}
\ead{giovannivincenzo.donatiello@unisalento.it}

\author[aff1,aff2,aff3]{Achille A. Nucita}
\ead{nucita@le.infn.it}

\author[aff1,aff2,aff3]{Francesco De Paolis}
\ead{depaolis@le.infn.it}

\author[aff1,aff2,aff3]{Antonio Franco}
\ead{franco@le.infn.it}

\author[aff3]{Francesco Strafella}
\ead{francesco.strafella@inaf.it}

\cortext[cor1]{Corresponding author}

\affiliation[aff1]{organization={Department of Mathematics and Physics ``E. De Giorgi'', Università del Salento},
            addressline={Via per Arnesano}, 
            city={Lecce},
            postcode={73100}, 
            country={Italy}}

\affiliation[aff2]{organization={INFN, Sezione di Lecce},
            addressline={Via per Arnesano}, 
            city={Lecce},
            postcode={73100}, 
            country={Italy}}

\affiliation[aff3]{organization={INAF, Sezione di Lecce},
            addressline={c/o Dipartimento Matematica e Fisica, Via per Arnesano}, 
            city={Lecce},
            postcode={73100}, 
            country={Italy}}

\begin{abstract}
We present LUMA, a convolutional neural network (CNN) pipeline for the automated detection of strong gravitational lenses in simulated astronomical imaging. The method combines a physically motivated preprocessing stage, which enhances faint arc and ring features, with a compact three-block CNN trained using class reweighting and modern learning-rate scheduling. On simulated data, the model reaches test accuracies of about $96$\% and receiver operating characteristic (ROC) area-under-the-curve (AUC) values of $\simeq 0.99$ for the non-trivial classes, while confusion-matrix analysis shows high completeness and purity for lens candidates. These results demonstrate that relatively lightweight CNN architectures can provide a competitive baseline for strong-lens searches, and they motivate future extensions toward real survey images and transformer-based models.
\end{abstract}


\end{frontmatter}

\section{Introduction}\label{sec:intro}
\textit{Strong gravitational lensing} occurs when a massive foreground object, such as a galaxy or a galaxy cluster, produces multiple, highly distorted and often strongly magnified images, including extended arcs or complete \textit{Einstein rings}, of a single background source, owing to the pronounced curvature of spacetime along the line of sight. Strong-lensing systems provide powerful and largely independent constraints on the total mass distribution of galaxies and clusters, including their dark matter component, and act as natural telescopes that enhance the observability of distant, high-redshift galaxies and active galactic nuclei. Furthermore, when combined with high-quality imaging and spectroscopic data, detailed lens modeling enables stringent tests of structure-formation scenarios and offers complementary probes of cosmological parameters, for example through time-delay measurements between multiple images of variable background sources.

Over the last decade, the identification of strong gravitational lenses has increasingly relied on \textit{machine-learning} (ML) techniques, in particular on \textit{convolutional neural networks} (CNNs), because of their ability to process the massive volumes of imaging data produced by current and forthcoming wide-field surveys. Traditional approaches based on visual inspection or on hand-crafted features are no longer viable at the data rates expected from facilities such as Euclid, from space, or the Vera C. Rubin Observatory, from the ground. As a consequence, \textit{deep-learning} (DL) and \textit{computer-vision} methods have become central components of the strong-lens detection strategies for large survey programs \citep{pearce2025}.

A number of studies have demonstrated the effectiveness of CNNs in recognizing lensing features in survey data and simulations. \citet{petrillo2017,petrillo2019} developed CNN classifiers applied to the \textit{Kilo-Degree Survey} (KiDS) and the \textit{Canada-France-Hawaii Telescope Legacy Survey} (CFHTLS), achieving high completeness and purity in the identification of strong lenses. \citet{jacobs2019} extended this work by training deeper architectures on increasingly realistic mock data. In preparation for Euclid, \citet{metcalf2019} organized the \textit{Strong Gravitational Lens Finding Challenge}, in which a wide range of CNN-based and other ML methods were benchmarked on their ability to recover strong lenses under realistic observing conditions. These efforts have laid the foundations for automated pipelines that will operate on the Euclid data stream.

In parallel with the development of detection pipelines, growing attention has been devoted to the automated estimation of lens parameters. \citet{hezaveh2017} pioneered the use of neural networks (NNs) for mass-model inference directly from lensed images, an approach subsequently refined and extended by \citet{morningstar2018} and \citet{schuldt2021}. Such automated inference tools are being actively explored for integration into the Euclid analysis framework, with the aim of enabling near real-time extraction of scientific information from newly detected lenses.

The present work builds upon these developments and is directly inspired by \citet{donatiello2025}, where a CNN tailored to the detection of \textit{low-surface-brightness} (LSB) galaxies achieved robust performance even at very faint flux levels. Motivated by these results, the goal of this paper is to assess whether a similar methodology can be successfully adapted to the identification of strong gravitational lenses in simulated data, and to characterize the regimes in which such an approach is expected to be most effective.

Rather than proposing a radically new architecture for strong-lens detection, the present work aims to provide a compact, well-understood baseline and a diagnostic framework for upcoming survey pipelines. LUMA combines a physically motivated preprocessing tailored to KiDS-like data with a lightweight three-block CNN whose behavior is analyzed both in image space and in lens–source parameter space via tabular machine-learning models. This design allows us to isolate the impact of preprocessing and lens-parameter dependence on performance, and to quantify in which regimes a relatively simple architecture is already sufficient and where more sophisticated models are likely to be needed.

\section{Dataset}\label{sec:data}
In this paper, we use the \textit{KiDS ESO-DR4} data release \citep{dejong2017}, which consists of galaxy images and multi-band catalogs for 292 survey tiles. After downloading the photometric catalogs, we select luminous red galaxies (LRGs; \citealt{eisenstein2001}), which are expected to be massive and therefore more likely to act as strong lenses for background sources.

Following \citet{petrillo2017}, LRGs are selected by requiring that $z < 0.4$, $r < 20$, $\lvert c_{\text{perp}} \rvert < 0.2$, and
$r < 14 + c_{\text{par}}/0.3$, where
\begin{equation}
c_{\text{par}} = 0.7\,(g - r) + 1.2\,\big[(r - i) - 0.18\big]
\end{equation}
and
\begin{equation}
c_{\text{perp}} = (r - i) - \frac{(g - r)}{4.0} - 0.18\,.
\end{equation}
In this scheme, $z$ denotes the photometric redshift, $r$ is the apparent magnitude in the $r$ band, and $g$, $r$, and $i$ are the magnitudes in the corresponding filters, so that $(g-r)$ and $(r-i)$ represent galaxy colors. The quantities $c_{\text{par}}$ and $c_{\text{perp}}$ are the “parallel” and “perpendicular” color combinations used in the SDSS LRG selection \citep{eisenstein2001}, which isolate luminous red galaxies by selecting objects lying close to the red-sequence ridgeline in color–color space. With this selection, the resulting sample contains about $85{,}000$ candidates, for each of which we extract the associated image as a $101 \times 101$ pixel cutout centered on the target.

We then simulate $85{,}000$ strong-lensing images with the same image size and pixel scale ($0.21~\mathrm{arcsec}\,\mathrm{px}^{-1}$) as the KiDS images. The simulations are performed by modeling the source galaxy with a S\'ersic profile with index $n$ in the range $0.5$--$5.0$, and by drawing the other lens and source parameters uniformly from the ranges listed in Table~\ref{table_sim}. With this setup, we are able to produce a variety of strong-lensing morphologies (rings, arcs, quads, folds, and cusps), as described in \citet{Schneider1992}.

\begin{table}[t]
\centering
\caption{Ranges of the lens and source parameters used in the mock-lens simulations.}
\label{table_sim}
\begin{tabular}{lcc}
\hline
Source parameter        & Range    & Units  \\
\hline
Einstein radius         & 1.4--5.0 & \arcsec \\
Axis ratio              & 0.3--1.0 & --     \\
Major-axis angle        & 0--180   & deg    \\
External shear          & 0.0--0.05& --     \\
External-shear angle    & 0.0--180 & deg    \\
\hline
Lens parameter          & Range    & Units  \\
\hline
Effective radius        & 0.2--0.6 & \arcsec \\
Axis ratio              & 0.3--1.0 & --     \\
Major-axis angle        & 0.0--180 & deg    \\
S\'ersic index          & 0.5--5.0 & --     \\
\hline
\end{tabular}
\end{table}

From this sample, we construct a mock dataset by randomly pairing each KiDS galaxy with a simulated strong-lensing image and combining them into composite cutouts. In practice, we generate $N_{\text{lens}} = 15{,}000$ lens systems (KiDS galaxy + simulated lens) and $N_{\text{non-lens}} = 15{,}000$ non-lensed galaxies (KiDS only), for a total of $N_{\text{tot}} = 30{,}000$ mock images (\ref{fig:sample}).

\begin{figure}
\centering
\includegraphics[width=0.8\linewidth]{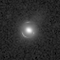}
\caption{Example of a simulated galaxy--galaxy strong lens from the mock dataset.}
\label{fig:lens}
\end{figure}

During mock-dataset production, we augment the data by applying the following transformations to both the galaxy and lens images:
\begin{enumerate}
  \item a random rotation between 0 and $2\pi$;
  \item a random shift in both $x$ and $y$ directions between $-4$ and $+4$ pixels;
  \item a 50\% probability of a horizontal flip;
  \item a rescaling of the lens image such that its maximum brightness matches the maximum brightness of the associated galaxy, multiplied by a uniform random factor in the range $0.02$--$0.2$.
\end{enumerate}
Each composite image is then cropped to a $60 \times 60$ pixel cutout to remove blank pixels introduced by the rotation and is processed to enhance faint structures associated with the lens features. This is achieved by applying a square-root stretch and a 96\% percentile cut on the pixel-brightness distribution. Finally, each image is converted to a gray-level representation by linearly rescaling the pixel values to the range 0--255.

The final mock dataset thus contains $N_{\text{tot}} = 30{,}000$ images, of which $N_{\text{lens}} = 15{,}000$ are lens systems and $N_{\text{non-lens}} = 15{,}000$ are non-lensed galaxies. We randomly split this dataset into a class-balanced training set of $N_{\text{train}} = 18{,}000$ images and a held-out test set of $N_{\text{test}} = 12{,}000$ images. All performance metrics and diagnostic plots reported in the following sections refer to this held-out test set, unless otherwise specified.

In addition, for the ablation study described in Sect.~\ref{subsec:ablation}, we construct an independent mock test set of $8{,}695$ images, consisting of $4{,}308$ lens systems and $4{,}387$ non-lens galaxies. This second test set is not used during training and provides a representative, nearly balanced sample for quantifying the impact of different preprocessing and data-augmentation choices on classifier performance.

Although the simulations are matched to the KiDS pixel scale and approximate seeing, we emphasize that they do not yet capture the full complexity of real survey images. In particular, we do not model spatial PSF variations across the field, diffraction spikes from bright stars, cosmic rays, or complex blends with unrelated foreground and background objects. The performance achieved on the simulated dataset should therefore be regarded as an optimistic upper limit, and the additional tests on real KiDS cutouts in Sect.~\ref{subsec:kids} are essential to assess how well the model transfers to the observational domain.

\section{Pipeline description}\label{sec:pipeline}
\subsection{Image preprocessing and normalization}\label{subsec:preprocessing}

The image preprocessing pipeline is designed to convert the raw data into a standardized input suitable for CNN training, while preserving the faint structures that characterize lenses \citep{petrillo2017,petrillo2019,vago2023,more2024,pearce2025}. The dataset consists of a directory of FITS files, each containing a 2D image and a header that stores the object's class label. For every valid file, the main image is loaded as a floating-point array and the header is read to recover the class information. The image is then converted to 32-bit floats and any invalid or missing values are cleaned to prevent numerical problems in later processing steps.

Let the original image be denoted by \(I(x,y)\), where \(x\) and \(y\) indicate the pixel coordinates. The first major step is the enhancement of faint structures, implemented through a combination of percentile-based intensity scaling and a nonlinear transformation. Two percentile thresholds are defined, a lower percentile \(P_{\text{low}}\) and an upper percentile \(P_{\text{high}}\), computed from the distribution of pixel values in \(I\). In practice, \(P_{\text{low}}\) and \(P_{\text{high}}\) correspond to, for example, the 1st and 99th percentiles of the pixel intensities, derived as
\begin{equation}\label{eq:3.1.1}
    p_{\min} = \text{percentile}(I, P_{\text{low}}), \qquad
    p_{\max} = \text{percentile}(I, P_{\text{high}}).
\end{equation}
These values define a dynamic range that excludes extreme outliers or very low background noise \citep{petrillo2017,petrillo2019}.

Using \(p_{\min}\) and \(p_{\max}\), an intermediate normalized image \(I_{\text{norm}}(x,y)\) is constructed by clamping and rescaling the original intensities:
\begin{equation}\label{eq:3.1.2}
    I_{\text{norm}}(x,y) = \frac{\max\big( \min(I(x,y), p_{\max}), \, p_{\min} \big) - p_{\min}}{p_{\max} - p_{\min} + \epsilon},
\end{equation}
where \(\epsilon\) is a small constant added to avoid division by zero. This mapping sends intensities below \(p_{\min}\) to 0 and those above \(p_{\max}\) to 1, while linearly stretching intermediate values to the \([0,1]\) interval. The effect is to emphasize the central intensity range where most astrophysical signal resides, while reducing the influence of extreme pixel values.

On top of this linear scaling, a nonlinear enhancement is applied to boost the visibility of faint arcs and rings. A power-law (\textit{gamma}) transform is adopted, controlled by a parameter \(\gamma\):
\begin{equation}\label{eq:3.1.3}
    I_{\text{enh}}(x,y) = \big( I_{\text{norm}}(x,y) \big)^{\gamma}.
\end{equation}
For \(\gamma < 1\), this transformation brightens low-intensity pixels relative to high-intensity ones, effectively lifting faint structures out of the background while compressing very bright cores, as discussed by \citet{lupton2004}. As an alternative, a logarithmic transform can be used,
\begin{equation}\label{eq:3.1.4}
I_{\text{enh}}(x,y) = \frac{\log\big(1 + I_{\text{norm}}(x,y)\big)}{\log 2},
\end{equation}
which similarly enhances low values but with a different contrast behavior; in this pipeline, the power-law transform is the default choice.

After the intensity and contrast adjustments, a spatial smoothing step is carried out to reduce pixel-scale noise without erasing the extended shapes of gravitational arcs. This is achieved using either a median filter or a mean (uniform) filter, depending on a method flag. For the median filter, a square window of size \(k \times k\) is slid across the image, and each pixel is replaced by the median of the values within that window:
\begin{equation}\label{eq:3.1.5}
I_{\text{blur}}(x,y) = \text{median}\big\{ I_{\text{enh}}(u,v) \,\big|\, (u,v) \in \mathcal{N}_{k}(x,y) \big\},
\end{equation}
where \(\mathcal{N}_{k}(x,y)\) denotes the local neighborhood of size \(k \times k\) centered on \((x,y)\). Alternatively, a mean filter computes the average over the same neighborhood,
\begin{equation}\label{eq:3.1.6}
I_{\text{blur}}(x,y) = \frac{1}{\lvert \mathcal{N}_{k}(x,y) \rvert} \sum_{(u,v) \in \mathcal{N}_{k}(x,y)} I_{\text{enh}}(u,v),
\end{equation}
yielding a more uniformly smoothed image but with slightly more blurring of sharp edges \citep{petrillo2019,more2024}.

The resulting image \(I_{\text{blur}}(x,y)\) is then converted to an 8-bit representation to facilitate further processing and visualization. To do so, the blurred image is first renormalized to \([0,1]\) using its own minimum and maximum values:
\begin{equation}\label{eq:3.1.7}
I_{\text{scaled}}(x,y) = \frac{I_{\text{blur}}(x,y) - I_{\min}}{I_{\max} - I_{\min} + \epsilon},
\end{equation}
with
\begin{equation}\label{eq:3.1.8}
I_{\min} = \min_{x,y} I_{\text{blur}}(x,y), \qquad
I_{\max} = \max_{x,y} I_{\text{blur}}(x,y).
\end{equation}
This normalized image is then mapped to the 8-bit integer range \([0,255]\) via
\begin{equation}\label{eq:3.1.9}
I_{8\text{bit}}(x,y) = \text{round}\big( 255 \times I_{\text{scaled}}(x,y) \big),
\end{equation}
and, if needed for the CNN input, recast to a floating-point representation in \([0,1]\) by dividing by 255.

An important step is the enforcement of a fixed spatial size for all images. The CNN in this pipeline expects \(60 \times 60\) inputs, so any preprocessed image whose dimensions differ from this size is centrally cropped (or, in principle, padded) to match. Suppose the processed image has height \(H\) and width \(W\). If \(H > 60\) or \(W > 60\), the starting indices for cropping are chosen as
\begin{align}\label{eq:3.1.10}
y_{\text{start}} &= \max\left(0, \left\lfloor \frac{H - 60}{2} \right\rfloor \right), \\
x_{\text{start}} &= \max\left(0, \left\lfloor \frac{W - 60}{2} \right\rfloor \right),
\end{align}
and the central \(60 \times 60\) patch is extracted:
\begin{equation}\label{eq:3.1.11}
I_{\text{crop}}(x,y) = I_{8\text{bit}}(x_{\text{start}} + x, \; y_{\text{start}} + y),
\quad 0 \le x,y < 60.
\end{equation}
This central-cropping strategy ensures that the main object, which is typically located near the center of the field in survey cutouts, remains in the frame, while edge regions are discarded. For images that already have the correct size, this step simply preserves the original dimensions.

Once all images have been preprocessed and resized, they are stacked into a four-dimensional array of shape \((N, 1, 60, 60)\), where \(N\) is the number of samples and the final dimension represents the single grayscale channel. In parallel, the corresponding class labels are collected from the FITS headers and stored as an integer vector. At this point, global dataset statistics are computed: the mean \(\mu\) and standard deviation \(\sigma\) of pixel values across the entire training set are obtained via
\begin{align}\label{eq:3.1.12}
\mu &= \frac{1}{N \times 60 \times 60} \sum_{i=1}^{N} \sum_{x,y} I_i(x,y), \\
\sigma &= \sqrt{\frac{1}{N \times 60 \times 60} \sum_{i=1}^{N} \sum_{x,y} \big( I_i(x,y) - \mu \big)^2 }.
\end{align}
These statistics are later used to perform per-channel normalization of the input tensors during training and testing, i.e.\ subtracting \(\mu\) and dividing by \(\sigma\) so that the network sees inputs with approximately zero mean and unit variance \citep{vago2023,porter2023}.

\begin{figure}
    \centering
    \includegraphics[width=0.6\linewidth]{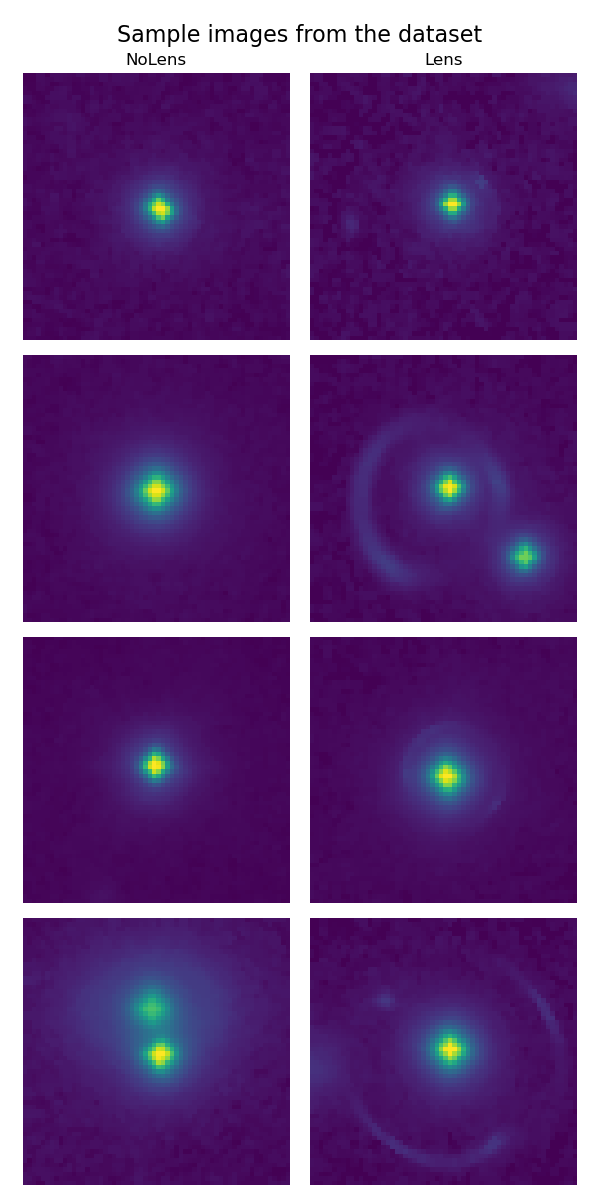}
    \caption{Sample preprocessed images from the simulated dataset, showing examples from both the non-lens and lens classes.}
    \label{fig:sample}
\end{figure}

To validate the effectiveness of the preprocessing pipeline qualitatively, a small subset of images from each class is displayed (Figure~\ref{fig:sample}). For each class label present in the dataset, a few images are randomly selected and arranged in a grid, with one column per class and several rows of examples. This visual inspection step helps confirm that the chosen percentile thresholds, gamma value, and smoothing kernel successfully enhance gravitational-lens features, such as arcs or rings, without introducing artifacts or erasing relevant structure. It also allows a quick check that all images share the same scale and dynamic range.

\subsection{CNN architecture}\label{subsec:architecture}
\begin{figure*}
    \centering
    \includegraphics[width=0.7\linewidth]{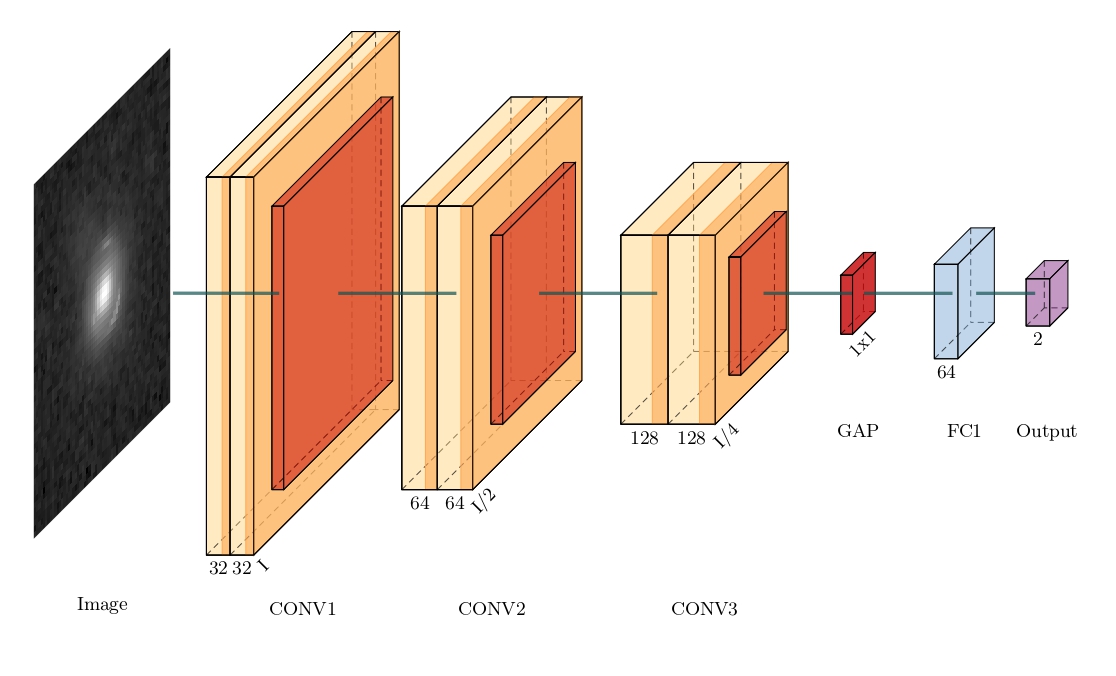}
    \caption{Diagram of the CNN architecture. The diagram is color-coded as follows: the input image in dark gray, convolutional layers in light orange, batch-normalization layers in a darker orange, the global average pooling (GAP) layer in red, and the fully connected and output layers in blue and violet, respectively. The thick lines represent the connections between these layers. The thickness of each block is proportional to the depth of the layer in the network, representing either the number of feature maps (in convolutional layers) or the number of neurons (in fully connected layers). Each convolutional block includes two convolutional layers, batch normalization, and a ReLU activation. In particular, the CONV1 block has a depth of 32, indicating 32 filters, CONV2 has a depth of 64, and CONV3 has a depth of 128, while FC1 has a depth of 64 neurons and the final output layer has 2 neurons. The notations such as $I$, $I/2$, and $I/4$ in the diagram indicate the input size at each stage of the network: $I$ represents the initial input size (e.g. $60 \times 60$), while $I/2$ and $I/4$ denote that the spatial dimensions (height and width) of the feature maps have been reduced by factors of 2 and 4, respectively, due to pooling or strided operations. The same convention applies to the other blocks. The figure is based on \textit{PlotNeuralNet} \citep{iqbal2018}.}
    \label{fig:scheme}
\end{figure*}

The CNN adopted in this work, hereafter referred to as LUMA (\textit{LUminosity Machine-learning Algorithm}), is structured as a sequence of \textit{convolutional blocks} followed by a \textit{classifier} head. The overall design (Figure~\ref{fig:scheme}) follows the common paradigm of progressively increasing the number of feature channels while reducing the spatial resolution, allowing the network to extract increasingly abstract representations of the input images, from simple edges to complex arc-like configurations typical of strong gravitational lenses \citep{petrillo2017,petrillo2019,vago2023,more2024}. The network takes as input a single-channel image \(I \in \mathbb{R}^{1 \times 60 \times 60}\) (in channels-first convention) and outputs a two-dimensional probability vector corresponding to the non-lens and lens classes.

The architecture is composed of three convolutional blocks. Each block consists of two successive convolutional layers with \(3 \times 3\) kernels and stride 1, each followed by batch normalization and a rectified linear unit (ReLU) activation, a \(2 \times 2\) max-pooling layer that halves the spatial dimensions, and a dropout layer that randomly zeros a fraction of activations during training. In the first block, the input channel dimension is mapped from 1 to 32 feature maps. Denoting the input to the first convolution by \(X^{(0)} \in \mathbb{R}^{1 \times 60 \times 60}\), the first convolutional layer applies 32 filters \(\{W^{(1)}_k\}_{k=1}^{32}\), each of size \(3 \times 3\), to produce pre-activation feature maps
\begin{equation}\label{eq:3.2.1}
Z^{(1)}_k = W^{(1)}_k * X^{(0)} + b^{(1)}_k,
\end{equation}
where \(*\) denotes the 2D convolution and \(b^{(1)}_k\) is a scalar bias term. Batch normalization then rescales and recenters each channel,
\begin{equation}\label{eq:3.2.2}
\tilde{Z}^{(1)}_k = \gamma^{(1)}_k \frac{Z^{(1)}_k - \mu^{(1)}_k}{\sqrt{(\sigma^{(1)}_k)^2 + \epsilon}} + \beta^{(1)}_k,
\end{equation}
with learnable parameters \(\gamma^{(1)}_k\) and \(\beta^{(1)}_k\) and running estimates \(\mu^{(1)}_k\) and \(\sigma^{(1)}_k\), followed by the ReLU nonlinearity
\begin{equation}\label{eq:3.2.3}
X^{(1)}_k = \max\big(0, \tilde{Z}^{(1)}_k \big).
\end{equation}

A second convolution–batchnorm–ReLU sequence maps \(32 \rightarrow 32\) channels, preserving the spatial resolution due to unit stride and padding of one pixel. Subsequently, a \(2 \times 2\) max-pooling operation with stride 2 reduces the spatial size from \(60 \times 60\) to \(30 \times 30\), and a dropout layer with drop probability \(p = 0.3\) is applied to the resulting activations, randomly setting a fraction of elements to zero during training to reduce overfitting:
\begin{equation}\label{eq:3.2.4}
X^{(1)}_{\text{pool}} = \text{MaxPool}_{2 \times 2}(X^{(1)}), \qquad
X^{(1)}_{\text{drop}} = M^{(1)} \odot X^{(1)}_{\text{pool}},
\end{equation}
where \(M^{(1)}\) is a Bernoulli mask with parameter \(1-p\) and \(\odot\) denotes element-wise multiplication. This first block extracts low-level features such as edges and small-scale gradients, which have been shown to be effective building blocks in lens-finding CNNs \citep{petrillo2017,petrillo2019}.

The second convolutional block repeats the same structure but increases the number of channels from 32 to 64. The input to this block, \(X^{(2)} \in \mathbb{R}^{32 \times 30 \times 30}\), is processed by two \(3 \times 3\) convolutional layers (with appropriate padding), each followed by batch normalization and ReLU, to yield 64-channel feature maps. Max pooling once again halves the spatial dimensions, resulting in feature maps of size \(64 \times 15 \times 15\). The dropout probability in this block is increased to \(p = 0.4\), providing stronger regularization as the representational capacity grows. At this stage, the network learns mid-level patterns that combine edges and blobs into more complex morphological cues, such as partial arcs or multiple-image configurations, which are particularly relevant for lens detection \citep{vago2023,more2024}.

The third convolutional block further expands the channel dimension from 64 to 128. The input tensor \(X^{(3)} \in \mathbb{R}^{64 \times 15 \times 15}\) is again passed through two \(3 \times 3\) convolutional layers with batch normalization and ReLU, now producing 128 feature maps. A final max-pooling stage reduces the spatial resolution to approximately \(7 \times 7\), yielding an activation tensor \(X^{(3)}_{\text{pool}} \in \mathbb{R}^{128 \times H' \times W'}\) with \(H', W' \approx 7\). The dropout probability is set to \(p = 0.5\) in this last convolutional block, providing strong regularization on the highest-level features. These deep activations encode structures that can correspond to full or partial Einstein rings, multiple-image systems, or characteristic lens–source configurations, akin to what has been presented in other networks trained on strong-lens datasets \citep{petrillo2019,pearce2025}.

To reduce the dependence on the exact spatial size and to produce a compact descriptor, the network applies global average pooling through an adaptive average-pooling layer, following the approach introduced by \citet{lin2014}, which maps each of the 128 feature maps to a single scalar. Formally, for each channel \(k\),
\begin{equation}\label{eq:3.2.5}
h_k = \frac{1}{H' W'} \sum_{x=1}^{H'} \sum_{y=1}^{W'} X^{(3)}_{\text{pool},k}(x,y),
\end{equation}
producing a 128-dimensional feature vector
\[
\mathbf{h} = (h_1, \dots, h_{128})^\top \in \mathbb{R}^{128}.
\]
The resulting vector is then flattened and passed to the fully connected classifier head.

The classifier consists of a single fully connected hidden layer followed by the output layer. First, the pooled feature vector $\mathbf{h} \in \mathbb{R}^{128}$ is projected to a 64-dimensional hidden representation,
\begin{equation}\label{eq:3.2.6}
\mathbf{z}^{(\mathrm{fc1})} = W^{(\mathrm{fc1})} \mathbf{h} + \mathbf{b}^{(\mathrm{fc1})},
\end{equation}
where $W^{(\mathrm{fc1})} \in \mathbb{R}^{64 \times 128}$ and $\mathbf{b}^{(\mathrm{fc1})} \in \mathbb{R}^{64}$ are learnable parameters. A ReLU activation is applied to obtain
\begin{equation}\label{eq:3.2.7}
\mathbf{a}^{(\mathrm{fc1})} = \max\big(0, \mathbf{z}^{(\mathrm{fc1})} \big),
\end{equation}
and a dropout layer with $p = 0.5$ is used on $\mathbf{a}^{(\mathrm{fc1})}$ to further regularize the classifier:
\begin{equation}\label{eq:3.2.8}
\tilde{\mathbf{a}}^{(\mathrm{fc1})} = \mathbf{m}^{(\mathrm{fc1})} \odot \mathbf{a}^{(\mathrm{fc1})},
\end{equation}
with $\mathbf{m}^{(\mathrm{fc1})}$ a Bernoulli mask. Finally, a linear output layer maps this hidden representation to the two logits corresponding to the non-lens and lens classes,
\begin{equation}\label{eq:3.2.9}
\mathbf{z}^{(\mathrm{out})} = W^{(\mathrm{out})} \tilde{\mathbf{a}}^{(\mathrm{fc1})} + \mathbf{b}^{(\mathrm{out})},
\end{equation}
where $W^{(\mathrm{out})} \in \mathbb{R}^{2 \times 64}$ and $\mathbf{b}^{(\mathrm{out})} \in \mathbb{R}^{2}$.

Overall, LUMA provides sufficient capacity to learn complex lens morphologies while keeping the number of trainable parameters moderate. This design is consistent with architectures that have been successfully applied to strong-lens searches and related applications in the literature, and closely follows the philosophy adopted in \citet{donatiello2025}.

\subsection{Training and testing}\label{subsec:train}
The training and testing phases represent the central part of the classification pipeline. In this work, the dataset is randomly split into a training set and a held-out test set, containing exactly $N_{\text{train}} = 18{,}000$ and $N_{\text{test}} = 12{,}000$ images, respectively, which correspond to approximately 60\% and 40\% of the total sample, while preserving the class distribution in both subsets.

A first step consists in computing class weights to counteract the imbalance typical of astronomical datasets, where confirmed lenses constitute a tiny minority compared to non-lens galaxies and artifacts \citep{petrillo2017}. These weights follow the inverse-frequency prescription
\begin{equation}\label{eq:3.3.1}
w_c = \frac{N}{K \, n_c},
\end{equation}
where $N$ is the total number of images, $K$ the number of classes (here $K=2$), and $n_c$ the number of samples in class $c$. This ensures that the loss function assigns higher penalties to misclassifying rare lenses, so that the optimizer pays more attention to them during learning. The optimizer chosen is AdamW, initialized with a base learning rate $\eta_0 = 7 \times 10^{-4}$ and a weight decay of $10^{-4}$. AdamW applies decoupled $L_2$ regularization directly on the parameters after the Adam update, which typically results in better generalization than standard Adam with coupled weight decay, especially in deep CNNs \citep{smith2018}.

The training loop spans at most 80 epochs, with early stopping configured to interrupt training if the validation loss does not improve for 7 consecutive epochs, thus preventing overfitting. At the beginning of each epoch, the network is switched to training mode, enabling dropout layers and updating batch-normalization statistics. The training set is then traversed in mini-batches of 64 images. For each batch, a forward pass produces raw logits $\mathbf{z}$ for each class, which are converted into probabilities via a softmax \citep{paszke2019},
\begin{equation}\label{eq:3.3.2}
\hat{y}_c = \frac{\exp(z_c)}{\sum_{j=1}^{K} \exp(z_j)}\,.
\end{equation}
The training objective is a weighted cross-entropy loss,
\begin{equation}\label{eq:3.3.3}
\mathcal{L} = - \sum_{c=1}^{K} w_c \, y_c \, \log \hat{y}_c,
\end{equation}
where $y_c$ is the one-hot encoded true label for class $c$ and $w_c$ is the corresponding class weight. This formulation makes misclassification of lens images (which have larger $w_c$) contribute more strongly to the loss, a strategy that has proved effective in imbalanced lens-finding applications \citep{davies2019,more2024}.

After computing $\mathcal{L}$ for a batch, backpropagation is applied to obtain gradients with respect to all model parameters. Before invoking the optimizer step, the gradients are clipped to have a maximum $L_2$ norm of 1.0, i.e.
\begin{equation}\label{eq:3.3.4}
g_{\text{clipped}} = g \cdot \min\left(1,\; \frac{\tau}{\lVert g \rVert_2}\right),
\end{equation}
with $\tau = 1.0$, in order to prevent gradient explosion and stabilize training of the deeper convolutional blocks \citep{smith2018}. The AdamW update is then performed, using running estimates of the first and second moments of the gradients together with the decoupled weight-decay term. During the epoch, the training loss is accumulated and averaged over all batches, yielding an epoch-level training loss that typically decreases from values around $1$ toward $0.2$–$0.3$ as the CNN starts recognizing lens-like features more reliably.

A one-cycle learning-rate policy is adopted, in which the learning rate is first increased and then annealed, in line with the cyclical learning-rate strategy proposed by \citet{smith2017} and the disciplined hyperparameter-tuning framework of \citet{smith2018}. This is implemented through a scheduler that varies $\eta_t$ over the entire training run comprising $S$ update steps (epochs times batches per epoch) \citep{paszke2019,petrillo2017}. The learning rate starts at a small value $\eta_{\min}$, increases linearly up to a maximum $\eta_{\max} = 0.02$ over an initial warm-up phase, and then decreases back below the initial value, following approximately a triangular or cosine-shaped schedule. A simple piecewise-linear variant can be written as
\begin{equation}\label{eq:3.3.5}
\eta_t =
\begin{cases}
\eta_{\min} + \dfrac{t}{S_1}(\eta_{\max} - \eta_{\min}), & 0 \le t \le S_1, \\[6pt]
\eta_{\max} - \dfrac{t - S_1}{S_2}(\eta_{\max} - \eta_{\text{final}}), & S_1 < t \le S_1 + S_2,
\end{cases}
\end{equation}
where $t$ is the step index, $S_1$ and $S_2$ are the lengths of the increasing and decreasing phases, and $\eta_{\text{final}}$ is a small final learning rate on the order of $10^{-6}$ \citep{donatiello2025,paszke2019}. This policy allows the optimizer to quickly explore parameter space early on, then perform fine-grained updates near the end of training. In addition, a Reduce-on-Plateau scheduler monitors the validation loss on an epoch basis. If the validation loss does not improve for a given patience window, the current learning rate is multiplied by a factor smaller than one, typically 0.5,
\begin{equation}\label{eq:3.3.6}
\eta_{\text{new}} = \eta_{\text{old}} \times 0.5,
\end{equation}
down to a minimum threshold (e.g. $10^{-6}$) \citep{paszke2019}. This mechanism provides an additional safety net to reduce the learning rate when the model reaches a plateau, thereby improving convergence.

At the end of each epoch, a full validation pass is performed on the held-out test set. During this phase, the model is switched to evaluation mode, disabling dropout and using frozen batch-normalization statistics. The test set is processed in batches, computing the same cross-entropy loss (now unweighted for reporting) and aggregating predictions and softmax probabilities. The average validation loss and overall accuracy are recorded, and if the validation loss is lower than any previously observed value, the current model parameters are saved to disk as the “best” checkpoint. Early-stopping logic keeps track of how many epochs have elapsed since the last improvement; if this number exceeds the patience value, training is terminated and the best checkpoint is retained for subsequent testing and analysis.

\begin{figure*}
    \centering
    \includegraphics[width=0.45\linewidth]{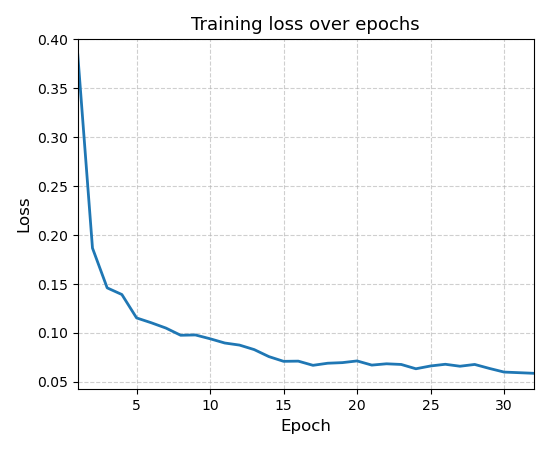}
    \includegraphics[width=0.45\linewidth]{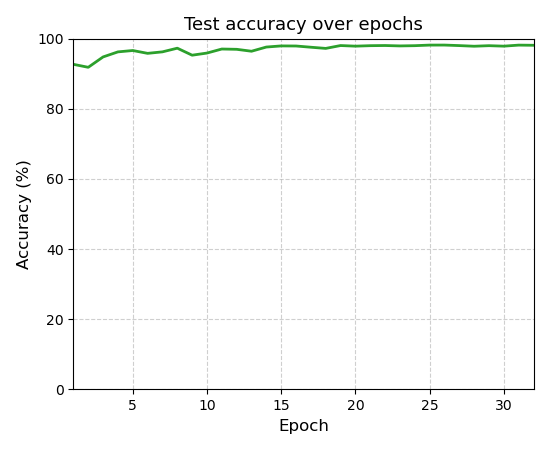}
    \caption{Evolution of training loss and test accuracy as a function of epoch for the adopted CNN. The loss decreases smoothly while the accuracy saturates after a few epochs, indicating stable convergence without signs of severe overfitting.}
    \label{fig:training_loss}
\end{figure*}

The testing stage, in the strict sense, uses this best saved model and focuses on a detailed assessment of performance on the held-out data (or on an independent dataset). The model remains in evaluation mode throughout testing. For each test batch, the CNN outputs logits and corresponding softmax probabilities $\hat{y}_c$. Predictions are obtained via $\hat{k} = \arg\max_c \hat{y}_c$, and these are compared to the true labels to compute the number of correct classifications and thus the overall accuracy,
\begin{equation}\label{eq:3.3.7}
\text{Accuracy} = \frac{\#\,\text{correct predictions}}{\#\,\text{total samples}}.
\end{equation}
In addition to accuracy and mean loss, the full arrays of true labels, predicted labels, and probabilities are stored to build more advanced diagnostics.

From these outputs, a confusion matrix (CM) is constructed, where rows correspond to true classes and columns to predicted classes. This matrix highlights, in particular, the rate at which real lenses (class 2) are misclassified as non-lenses, which is crucial for understanding completeness and purity in survey applications. Receiver operating characteristic (ROC) curves are derived by sweeping a probability threshold and computing the true positive rate (TPR),
\begin{equation}\label{eq:3.3.8}
\text{TPR} = \frac{TP}{TP + FN},
\end{equation}
and the false positive rate (FPR),
\begin{equation}\label{eq:3.3.9}
\text{FPR} = \frac{FP}{FP + TN},
\end{equation}
at each threshold. The area under the ROC curve (AUC) provides a threshold-independent measure of separability, with values close to 1 indicating excellent discrimination between lenses and non-lenses \citep{pearce2025}.

\section{Results}\label{sec:result}
The performance of the proposed pipeline can be assessed at three complementary levels: global convergence of the CNN during training, classification quality on the held-out test set, and interpretability of the model behavior through lens-parameter analysis. Taken together, these results indicate that the network learns a stable and highly discriminative representation of the simulated strong-lens population, in line with what is reported for similar CNN-based lens finders in the literature \citep{petrillo2017,davies2019,vago2023,more2024,pearce2025}.

\begin{figure}
    \centering
    \includegraphics[width=0.7\linewidth]{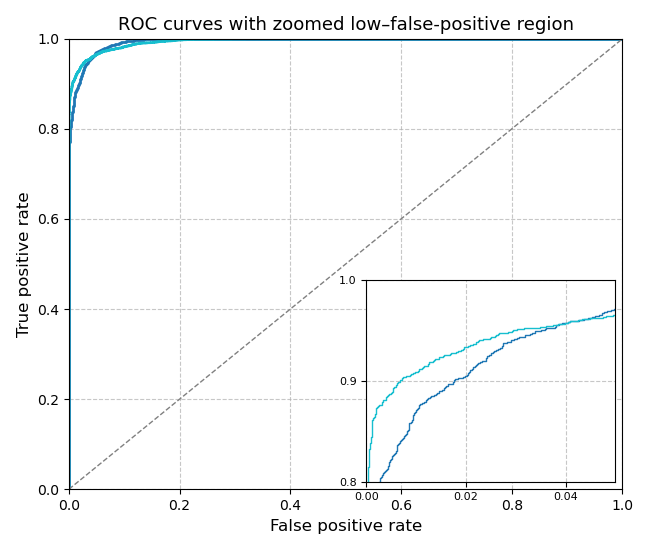}
    \caption{ROC curves for the lens vs.\ non-lens classification on the held-out test set. The panel includes an inset zooming into the low–false-positive region, where the classifier is expected to operate, and shows areas under the curve close to unity, indicating excellent discriminative performance.}
    \label{fig:roc}
\end{figure}

\begin{figure}
    \centering
    \includegraphics[width=0.7\linewidth]{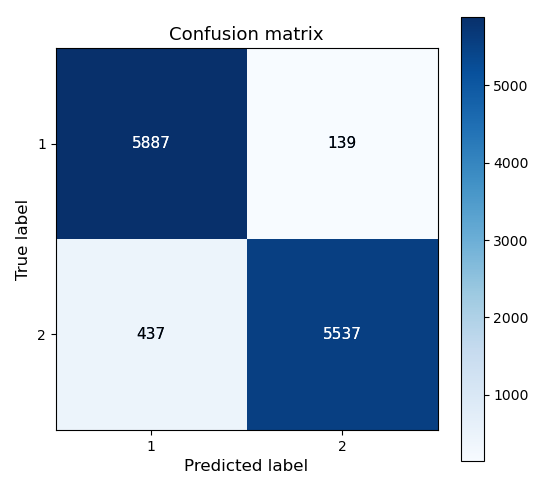}
    \caption{Confusion matrix for the lens vs.\ non-lens classification on the test set ($N_{\text{test}} = 12{,}000$ images). The color scale encodes the number of objects in each cell, with darker shades corresponding to higher counts along the diagonal (correct classifications) and lighter shades indicating fewer misclassified examples in the off-diagonal entries.}
    \label{fig:cm}
\end{figure}

From an optimization perspective, LUMA exhibits rapid and smooth convergence (Figure~\ref{fig:training_loss}). The training loss decreases monotonically from an initial value of $\simeq 0.5$ to about $0.06$ over 30 epochs, with the steepest decline occurring within the first ten epochs, after which the curve gradually flattens and approaches a stable plateau \citep{paszke2019}. In parallel, the test accuracy rises sharply from $\sim 90$\% in the first epoch to above 95\% after only a few epochs and then slowly increases, stabilizing around 96\% on the held-out test set without exhibiting the characteristic divergence between training and test metrics that would signal overfitting \citep{smith2018,monigatti2022}. Consistently with the confusion matrix in Figure~\ref{fig:cm}, the recall for the lens class is $\sim 93$\% and the recall for non-lenses is $\sim 98$\%, indicating that the network retains most genuine lenses while keeping the contamination by non-lenses relatively low. This behavior is consistent with an optimization regime in which the one-cycle learning-rate schedule and early-stopping criterion guide the network toward a well-generalizing minimum in parameter space, rather than memorizing the training set, as advocated in recent work on deep learning for astronomical imaging \citep{smith2018,monigatti2022}.

Representative cutouts from the ``NoLens'' and ``Lens'' classes show that non-lens examples are typically dominated by a single, compact galaxy with nearly circular isophotes, while lens images contain additional extended features such as partial or complete Einstein rings and multiple source images surrounding the central deflector \citep{petrillo2017,petrillo2019}. The preprocessing pipeline enhances these low-brightness structures, so that in the lens class the ring and arc components are clearly visible, whereas in the non-lens class the residual background appears largely featureless, in agreement with previous simulation-based lens-finding studies \citep{davies2019,vago2023}. This visual distinction suggests that the CNN is exposed to well-separated morphological manifolds in input space, which explains the fast separation of classes during training.

On the held-out test set, the discriminative power of the CNN is best quantified through ROC analysis. The ROC curves (Figure~\ref{fig:roc}) for the two non-trivial classes (corresponding to lens candidates versus contaminants) lie very close to the upper-left corner of the diagram across the full range of decision thresholds, with AUC values of $\simeq 0.99$ \citep{pearce2025,davies2019}. This indicates that, when samples are ranked by the predicted probability of belonging to the lens class, the model almost always assigns higher scores to true lenses than to non-lenses, a behavior that is particularly desirable for large surveys such as KiDS, DES, and Euclid, where high completeness at fixed false-positive rate is required \citep{petrillo2017,pearce2025}.

The confusion matrix offers a complementary, threshold-dependent view (Figure~\ref{fig:cm}). At a standard decision threshold of $0.5$, the matrix is dominated by large diagonal entries, with 5{,}887 correctly identified non-lenses and 5{,}537 correctly identified lenses, and comparatively small off-diagonal counts of 139 false positives and 437 false negatives. Only a small fraction of genuine lenses ($\sim 7$\%) are therefore missed, and the contamination of the predicted lens sample by non-lenses remains low, which is an attractive operating point for subsequent human or automated vetting, consistent with other CNN-based searches \citep{petrillo2019,more2024}.

\subsection{Interpretation of results}\label{subsec:ml}
\begin{figure*}
    \centering
    \includegraphics[width=0.7\linewidth]{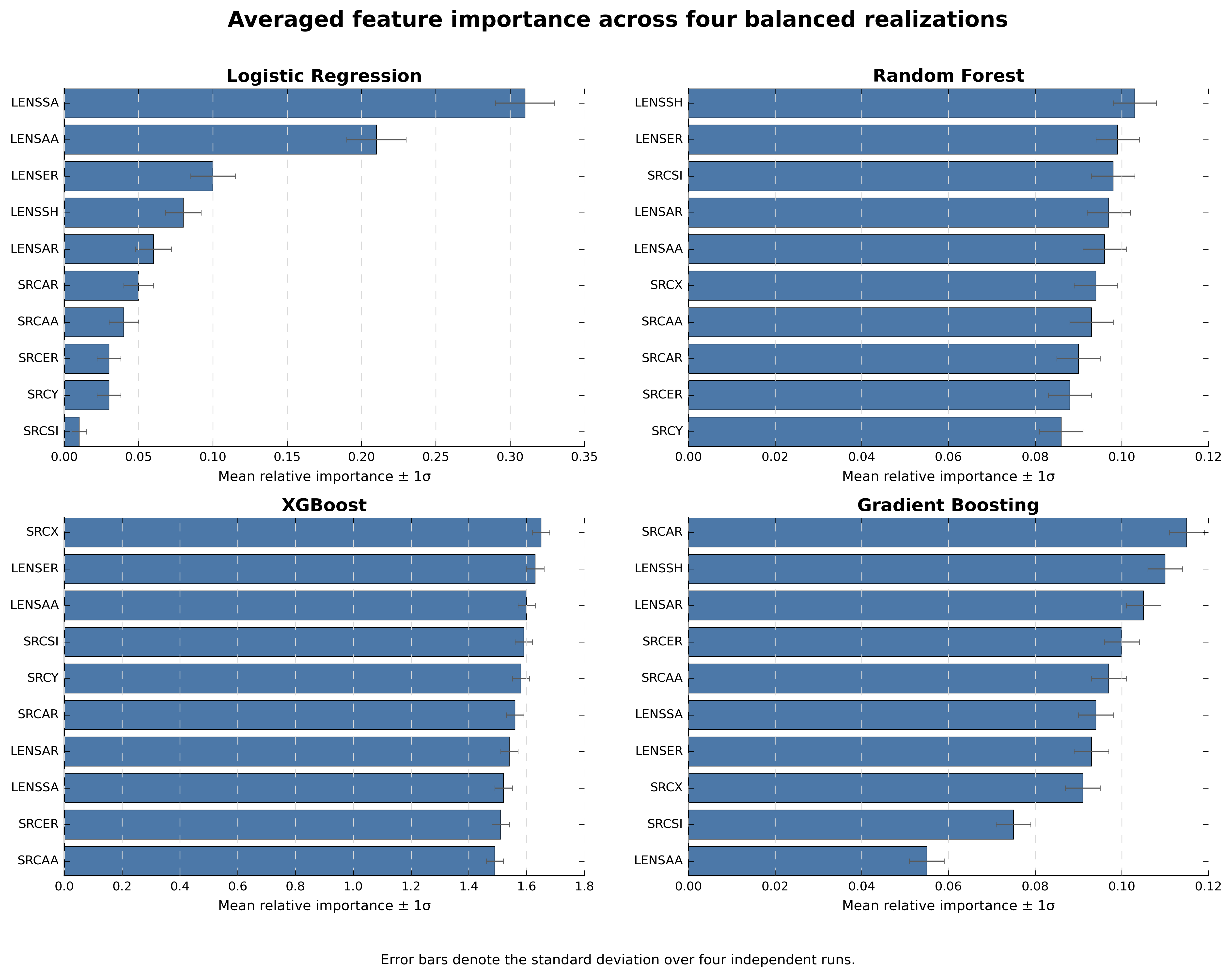}
    \caption{Mean feature importance across four balanced realizations for each classical model. Error bars denote the standard deviation over the four independent runs.}
    \label{fig:model_results}
\end{figure*}

To probe in more detail which physical properties of the lens–source system drive the success or failure of the CNN, we construct a tabular dataset of class-2 (strong-lens) examples in which each row is labeled as ``correct'' or ``incorrect'' according to the CNN prediction and is described by a set of scalar parameters extracted from the FITS headers (Einstein radius, lens and source axis ratios and position angles, source positions, and shapelet coefficients). On this dataset, balanced by undersampling correctly classified systems, we train a suite of classical machine-learning models to predict the correctness flag (Table~\ref{table_model_performance}), following approaches similar to those used in recent interpretability studies of deep networks in astronomy \citep{vago2023,more2024}. The corresponding mean feature-importance patterns across the four realizations are shown in Figure~\ref{fig:model_results}.

Linear logistic regression attains only modest performance, with accuracies of 49--54\% and F1 scores in the range 0.58--0.66, indicating that a purely linear decision boundary in this parameter space is insufficient to capture the complex dependencies underlying the CNN's behavior. In contrast, nonlinear models perform substantially better. Random forests achieve accuracies of $\sim 75$\% and F1 scores of $\sim 0.86$, gradient-boosting models and their histogram variant yield 69--73\% accuracy with F1 scores between 0.81 and 0.84, and $k$-nearest neighbors reaches a similar regime, with accuracies of $\sim 71$--73\% and F1 scores of $\sim 0.83$--0.84 \citep{Breiman2001,Friedman2001}. XGBoost shows intermediate performance, with accuracies of 69--70\% and F1 scores around 0.82 \citep{Chen2016}. These values imply that, given only the lens parameters, one can already predict with reasonably high confidence whether the CNN will classify a system correctly, reinforcing the idea that the network's errors are not random but instead correlated with specific regions of the lens-parameter space.

\begin{table*}[t]
\centering
\caption{Mean performance and run-to-run variability of classical models trained on balanced lens-parameter datasets across four independent realizations. The reported uncertainty corresponds to the standard deviation across runs.}
\label{table_model_performance}
\begin{tabular}{lcc}
\hline
Model                        & Accuracy mean $\pm \sigma$ [\%] & F1 mean $\pm \sigma$ \\
\hline
Logistic Regression          & $53.3 \pm 2.9$   & $0.64 \pm 0.04$ \\
Random Forest                & $74.9 \pm 0.4$   & $0.856 \pm 0.003$ \\
XGBoost                      & $70.2 \pm 1.7$   & $0.821 \pm 0.012$ \\
Gradient Boosting            & $73.2 \pm 1.6$   & $0.843 \pm 0.010$ \\
Histogram Gradient Boosting  & $70.3 \pm 1.5$   & $0.821 \pm 0.011$ \\
SVC                          & $55.5 \pm 2.2$   & $0.676 \pm 0.018$ \\
KNN                          & $72.1 \pm 0.8$   & $0.835 \pm 0.006$ \\
\hline
\end{tabular}
\end{table*}

There is, however, non-negligible variability across different runs of the tabular models, both in the global metrics (accuracy, F1) and in the detailed feature-importance rankings, which is fully consistent with the way the dataset is constructed and with the stochastic nature of the algorithms. Between two balanced realizations of the ``correct vs.\ wrong'' sample, for instance, the accuracy of logistic regression ranges from roughly 49\% to 54\% and the corresponding F1 score from $\sim 0.58$ to $\sim 0.66$, while random forests remain around $\sim 75$\% accuracy and F1 $\sim 0.86$ but show slightly different orderings of the most important parameters \citep{vago2023,more2024}. A similar behavior is observed for XGBoost and gradient-boosting models: all of them maintain F1 scores in the $\sim 0.81$--0.84 range, yet they swap the relative ranking of physically related quantities such as the lens Einstein radius (LENSER), the lens and source position angles (LENSAA and SRCAA), and the source shapelet coefficient (SRCSI), with some runs emphasizing predominantly lens-side parameters (e.g. LEN\-SAA, the lens position angle; LENSAR, the lens axis ratio; LENSSH, the lens shear) and others giving more weight to source descriptors (SRCAR, the source axis ratio; SRCAA, the source position angle; SRCX/SRCY, the source centroid coordinates).

Part of this dispersion arises from the construction of the tabular dataset itself. To obtain a balanced problem, the number of correctly classified lenses is reduced by random undersampling, so each training run sees a slightly different subset of ``correct'' systems and, consequently, a different empirical distribution of the input parameters. Models that are sensitive to the exact covariance structure of the features, most notably logistic regression, which fits a single linear decision boundary, are particularly affected by these small changes and therefore exhibit larger run-to-run fluctuations in their coefficients and importances. In addition, many of the algorithms introduce randomness internally: random forests sample subsets of examples and features for each tree; gradient-boosting and XGBoost variants use stochastic subsampling of rows and columns; SVC and KNN depend on the specific train/validation split. Even with fixed hyperparameters, two independent trainings thus produce ensembles of trees and decision boundaries that are not identical, leading to slightly different importance vectors.

An additional contribution comes from the strong correlations among the physical parameters. Quantities such as the Einstein radius, the lens and source axis ratios, the lens and source position angles, and the source shapelet coefficients describe related aspects of the same lens–source configuration and are therefore far from independent. In the presence of such multicollinearity, tree-based methods tend to share importance across interchangeable features: in one realization, a split may preferentially occur on LENSER, in another on LENSAA or SRCAR, yielding different bar heights in the plots while encoding essentially the same physical information redistributed among correlated variables. Finally, the number of misclassified lenses is relatively small compared to the correctly classified ones, so the correct/wrong label itself is noisy: small fluctuations in the sampled training set or in the random seeds can move borderline systems across the decision boundary, slightly shifting precision, recall, and F1.

Taken together, these results show that LUMA's mistakes are not random but cluster in specific regions of lens-parameter space. The CNN is most likely to fail on systems with small Einstein radii, low ellipticity, and less favorable source–lens alignments, i.e. when the lensed features are intrinsically less conspicuous and more easily confused with isolated galaxies. By contrast, highly elliptical, well-aligned configurations are almost always classified correctly. This suggests that future improvements should focus on architectures and training schemes that explicitly target these hard regimes, for instance via targeted augmentation of small-radius lenses or multi-scale feature extractors.

\subsection{Ablation study}\label{subsec:ablation}
We carried out a specific ablation study on the simulated dataset to verify the impact of the main design choices in the LUMA pipeline (see Table~\ref{tab:ablation_full}). As anticipated in Sect.~\ref{sec:data}, all ablation experiments were evaluated on an independent test set of 8{,}695 mock images, consisting of 4{,}387 non-lens galaxies and 4{,}308 lens systems. This split was not used during training and was constructed to provide a representative and nearly balanced sample for quantifying the impact of each modeling choice. For each configuration, we repeated the training multiple times with different random initializations and batch orderings, and we report the mean values and run-to-run standard deviations ($\pm\sigma$) for each metric. These uncertainties encompass both the stochastic variability of the training process and the intrinsic statistical sampling limits of the test set, which impose a theoretical lower bound on precision of $\sigma_{\text{binomial}} = \sqrt{p(1-p)/N} \approx 0.10\%$--$0.20\%$ for sample sizes of $N \approx 4{,}300$ per class. In all tests we kept the CNN architecture, the train/test split, and the loss function fixed, and varied only the preprocessing or the on-the-fly data augmentation.

We first compared the baseline configuration, which includes the gamma–percentile enhancement, with a variant where the enhancement is made purely linear by setting $\gamma = 1$. Both models achieve the same ROC AUC for the lens class ($0.990 \pm 0.010$), showing that the global ranking of lens versus non-lens probabilities is essentially unchanged. However, the baseline configuration converges to a lower loss and yields a higher test accuracy ($98.44 \pm 0.12$\%) compared to the linear variant ($97.28 \pm 0.15$\%), together with a higher recall for the lens class ($96.80 \pm 0.18$\% versus $95.40 \pm 0.20$\%) and a slightly higher recall for non-lenses ($99.22 \pm 0.10$\% versus $98.86 \pm 0.12$\%). These differences remain larger than the estimated run-to-run scatter, indicating that the gamma correction is not strictly required to separate the classes, but it sharpens the decision boundary around the operating threshold and improves the recovery of faint lens systems without degrading the performance on non-lenses.

\begin{table*}[t]
\centering
\caption{Ablation study on the main preprocessing and training choices of the LUMA pipeline, evaluated on an independent simulated test set of 8{,}695 images (4{,}387 non-lenses and 4{,}308 lenses). Quoted uncertainties correspond to the standard deviation across independent training runs with different random seeds, encompassing stochastic optimization variability and statistical sampling limits ($\sigma_{\text{binomial}} \approx 0.1\%$--$0.2\%$).}
\label{tab:ablation_full}
\begin{tabular}{lcccc}
\hline
Configuration & Accuracy [\%] & ROC AUC (lens) & Lens recall [\%] & Non-lens recall [\%] \\
\hline
Full pipeline (baseline, median blur) & $98.44 \pm 0.12$ & $0.99 \pm 0.01$ & $96.80 \pm 0.18$ & $99.22 \pm 0.10$ \\
No gamma (linear scaling)             & $97.28 \pm 0.15$ & $0.99 \pm 0.01$ & $95.40 \pm 0.20$ & $98.86 \pm 0.12$ \\
No augmentation                       & $98.10 \pm 0.15$ & $0.99 \pm 0.01$ & $96.82 \pm 0.20$ & $99.36 \pm 0.12$ \\
Mean blur                             & $89.85 \pm 0.18$ & $0.99 \pm 0.01$ & $97.99 \pm 0.21$ & $81.84 \pm 0.56$ \\
\hline
\end{tabular}
\end{table*}

We then tested the impact of removing the on-the-fly geometric data augmentation, i.e.\ disabling the random horizontal flips and small rotations applied at training time. In this case, the performance remains very similar to the baseline, with an accuracy of $(98.10 \pm 0.15)$\%, a lens recall of $(96.82 \pm 0.20)$\%, a non-lens recall of $(99.36 \pm 0.12)$\%, and ROC AUC $(0.99 \pm 0.01)$, all compatible with the baseline within one standard deviation. This behavior is expected, since the simulated training set already includes substantial geometric variability introduced during the mock-image generation stage, where rotations and shifts are applied directly to the synthetic lens systems. For the present KiDS-like simulations, the PyTorch augmentation thus acts mainly as an additional regularization layer rather than as a critical source of diversity.

Finally, we investigated the effect of replacing the median filter in the preprocessing stage with a mean filter, while keeping all other settings unchanged. In this configuration, the classifier achieves a lens recall of $97.99 \pm 0.21$\%, but its overall test accuracy drops substantially to $89.85 \pm 0.18$\% because the non-lens recall degrades to $81.84 \pm 0.56$\%, corresponding to 772--821 misclassified non-lenses out of 4{,}387 across different runs. The mean filter smooths noise aggressively, causing faint background fluctuations to mimic arc-like structures and triggering a high rate of false positives. This confirms that the median filter in the baseline LUMA configuration is essential to maintain a low false-positive rate. For this reason, we retain the median filter in the baseline configuration, since it provides a more balanced trade-off between lens completeness and contamination, which is crucial when building clean candidate catalogs for follow-up.

In summary, the ablation analysis confirms that the adopted baseline is not simply the configuration with the highest global accuracy, but the one providing the most balanced behavior across classes. Gamma enhancement improves the effective separability near the decision threshold, while the extra training-time augmentation has only a limited impact in the presence of already diversified simulations. By contrast, replacing the median filter with a mean filter leads to an increase in false positives, despite the slightly higher lens recall. For this reason, the baseline preprocessing is retained in the final LUMA pipeline.

\subsection{Validation on KiDS survey data}\label{subsec:kids}
To assess whether the trends identified on simulated data transfer to the observational domain, we further evaluated the model on real KiDS images. This test provides a stricter validation of the method, since survey images include instrumental effects, noise, blending, and a broader range of galaxy morphologies than the simulations used for training. For this purpose, we used the \textit{KiDS DR4 high-quality strong-lens catalog} \citep{kuijken2019,petrillo2019,li2021,li2020}, which contains 268 visually selected strong-lens candidates. One object was excluded from our test sample due to unusable imaging, so the resulting evaluation set comprises 267 candidates and therefore does not provide a balanced lens-versus-non-lens sample.

\begin{figure}
    \centering
    \includegraphics[width=0.8\linewidth]{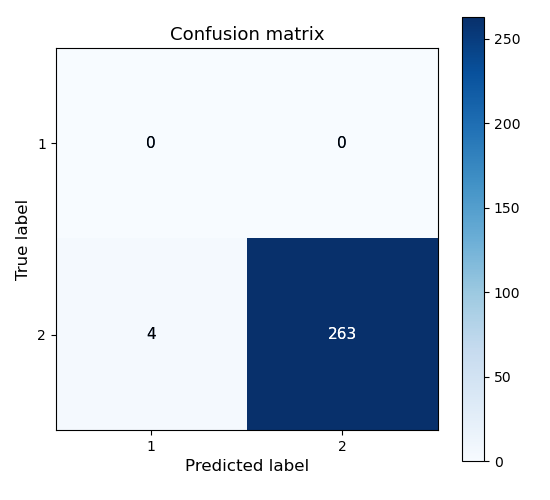}
    \caption{Confusion matrix for the test on the \textit{KiDS DR4 high-quality strong-lens catalog}.}
    \label{fig:kids}
\end{figure}

The corresponding confusion matrix (Figure~\ref{fig:kids}) shows that the classifier correctly identifies 263 lenses, misclassifying only 4 of them. This corresponds to a recall of approximately 98\% for the dominant class, indicating that the model recovers the main observational population with very high completeness. Taken together, these results suggest that the behavior observed on the real KiDS images is not only consistent with the simulation-based analysis, but also confirms that the model retains strong sensitivity to the primary strong-lensing morphology in survey data.

We also emphasize that, unlike the simulated dataset, this catalog does not provide a controlled non-lens population, so we only report completeness (recall) for real lenses and refrain from quoting overall accuracy or purity.

\section{Conclusions and future prospects}\label{sec:conclusion}
In this work we investigated the feasibility of using a CNN to identify simulated strong gravitational lenses in imaging data, and we then extended the analysis to real KiDS observations. Building on previous CNN-based lens-finding efforts \citep{petrillo2017,petrillo2019,davies2019,vago2023,more2024}, we designed a complete pipeline that combines a physically motivated image-preprocessing stage, a three-block CNN architecture tailored to $60 \times 60$ cutouts, and a dedicated training strategy based on class reweighting and modern learning-rate scheduling. On simulated data, the network achieves high test accuracy and ROC AUC values close to unity, indicating that it can robustly separate lenses from non-lenses under controlled conditions. Complementary tabular analyses based on lens-parameter headers further show that the CNN's successes and failures are strongly correlated with the underlying physical configuration, in particular with Einstein radius, ellipticity, and source–lens alignment. Despite its architectural simplicity, LUMA achieves ROC AUC $\simeq 0.99$ and competitive completeness and purity on KiDS-like simulations, while using an order of magnitude fewer parameters than many recent lens-finding CNNs \citep{davies2019}. This makes it attractive as a fast pre-filter in survey pipelines and as a controlled baseline to benchmark more complex models.

Importantly, the methodology was not limited to simulations: we also validated the pipeline on real KiDS images. This observational benchmark provides a more demanding test, since survey images include instrumental effects, noise, blending, and a broader range of galaxy morphologies than the simulations used for training. The corresponding confusion matrix shows that the classifier correctly identifies 263 class-2 systems, misclassifying only 4 of them, which translates into a recall of approximately 98\% for the dominant class. In this sense, the result is fully consistent with the composition of the KiDS sample and with the internal label structure adopted in this work: the classifier concentrates its predictions on the dominant class and reproduces the expected outcome for the observational test, while still demonstrating strong sensitivity to the primary strong-len\-sing morphology in real survey data.

At the same time, our results highlight some weaknesses. A key limitation of the present study is that the simulations, while KiDS-like, remain idealized in several respects and do not fully reproduce PSF variability, artifacts, and blending in the survey. The excellent ROC AUC obtained on simulations should thus be interpreted as a best-case scenario, whereas the KiDS tests provide a more realistic assessment of model robustness. Future work will focus on injecting simulated lenses into real KiDS cutouts and exploiting domain-adaptation techniques to bridge this gap more systematically. Even in the controlled setting of simulations, the confusion matrix reveals a small population of missed lenses, and the variability observed across different tabular models and resamplings indicates that the boundary between ``clear'' and ``ambiguous'' systems is not sharp, but rather shaped by correlations among multiple parameters and by the finite size of the training set. The real-data validation reinforces this point: although the model performs very well on the dominant observational class, directly extrapolating performance to more heterogeneous survey samples must still be done with caution, since real observations introduce additional complications such as PSF variability, background structures, blending, and other systematics that are only partially represented in the simulations.

Despite these caveats, the methodology developed here is flexible and readily extensible. LUMA can, with minor modifications, be applied to other problems in which faint or morphologically complex structures must be identified against a noisy background, such as the search for faint galaxies, as shown in \citet{donatiello2025}, planetary nebulae, stellar streams, open and globular clusters, and other rare transients. Looking forward, a natural next step would be to further expand the real-data component of the analysis by incorporating additional confirmed lenses and carefully selected hard negatives, while also exploring more sophisticated data-augmentation schemes and domain-adaptation techniques to narrow the gap between simulations and observations, along the lines proposed by \citet{davies2019} and \citet{vago2023}.

Looking ahead, there is substantial scope for exploring alternative architectures beyond classical CNNs. Recent work has demonstrated the potential of transformer-based vision models for strong-lens detection, such as the TEGLIE framework of \citet{grespan2024}, which employs vision transformer encoders on KiDS. Their ability to capture long-range correlations and global context may prove particularly advantageous for high-resolution space-based imaging, where lensing features can extend over large fractions of the field of view. Hybrid architectures that combine convolutional backbones with transformer blocks, or that jointly process imaging and tabular lens-parameter information, represent a promising avenue for future research.

Ultimately, this study reinforces the growing consensus that deep learning will play a central role in the exploitation of next-generation cosmological surveys. By demonstrating that the presented architecture can already achieve excellent perfor\-mance on simulated strong-lensing data, while also revealing where and why it fails, we provide a concrete baseline and a set of diagnostic tools upon which more advanced models can be built. Future work will focus on extending this framework to increasingly realistic data, integrating transformer-based architectures, and embedding the resulting models into end-to-end analysis pipelines, with the broader goal of enabling timely and reliable identification of the thousands of strong gravitational lenses expected in the next years.

\section*{Declaration of generative AI and AI-assisted technologies in the manuscript preparation process}
During the preparation of this work, the authors used Perplexity AI to improve the clarity and readability of the manuscript. After using this tool, the authors reviewed and edited the content as needed and take full responsibility for the content of the published article.

\section*{Declaration of competing interest}
The authors declare that they have no known competing financial interests or personal relationships that could have appeared to influence the work reported in this paper.

\section*{Acknowledgments}
We acknowledge support from the INFN projects TAsP and EUCLID. This work is partially supported by ICSC – Centro Nazionale di Ricerca in High Performance Computing.

Based on observations made with ESO Telescopes at the La Silla Paranal Observatory under program IDs 177.A-3016, 177.A-3017, 177.A-3018, and 179.A-2004, and on data products produced by the KiDS consortium. The KiDS production team acknowledges support from: Deutsche Forschungsgemein\-schaft, ERC, NOVA and NWO-M grants; Target; the University of Padova; and the University Federico II (Naples).

\section*{Data availability}
Data will be made available on request.

\bibliographystyle{elsarticle-harv}
\bibliography{reference}

\end{document}